\documentclass{svjour3}                     % onecolumn (standard format)
\smartqed  % flush right qed marks, e.g. at end of proof
\usepackage{graphicx}
\usepackage{mathptmx}      % use Times fonts if available on your TeX system
\usepackage{amsmath}
\journalname{J. Supercond. and Novel Magnetism}
\begin{document}

\title{Isotope Effect of Underdoped Cuprates in the Yang-Rice-Zhang Model
}
\titlerunning{YRZ Isotope Effect}        % if too long for running head

\author{E. Schachinger         \and
        J. P. Carbotte %etc.
}

\institute{E. Schachinger \at
              Institute of Theoretical and Computational Physics \\
              Graz University of Technology, NAWI Graz, A-8010 Graz, Austria\\
              \email{schachinger@itp.tu-graz.ac.at}
           \and
           J. P. Carbotte \at
           Department of Physics and Astronomy, McMaster University,\\ Hamilton Ont., Canada L8S 4M1\\
           The Canadian Institute for Advanced Research, Toronto Ont., Canada M5G 1Z8
}

\date{Received: date / Accepted: date}

\maketitle

\begin{abstract}
The underdoped region of the cuprates phase diagram displays many novel
electronic phenomena both in the normal and the superconducting state.
Many of these anomalous properties have found a natural explanation
within the resonating valence bond spin liquid phenomenological model
of Yang-Rice-Zhang (YRZ) which includes the rise of a pseudogap. This
leads to Fermi surface reconstruction and profoundly changes the electronic
structure. Here we extend previous work to consider the shift in critical
temperature on $^{16}$O to $^{18}$O substitution, The isotope effect has been
found experimentally to be very small at optimal doping yet to rapidly increase
to very large values with underdoping. The YRZ model provides a natural
explanation of this behavior and supports the idea of a pairing mechanism
which is mainly spin fluctuations with a subdominant $(\sim 10\%)$ phonon
contribution.
\keywords{Isotope Effect \and Underdoped Cuprates \and Yang-Rice-Zhang Model}
\PACS{74.20.Mn \and 74.62.Yb \and 74.72.Kf}
% \subclass{MSC code1 \and MSC code2 \and more}
\end{abstract}

\section{Introduction}
\label{intro}
The isotope effect ($\alpha$) which gives the change in critical
temperature with ion mass played a significant role in the development
of the BCS theory of conventional metals. It was widely interpreted
to indicate that the electron-phonon interaction was involved in the
mechanism of Cooper pair condensation. On the other hand, in the
high $T_c$ oxides at optimum doping $\alpha$ is an order of magnitude
smaller \cite{franck91,franck94} than the canonical BCS value of
$1/2$ and this was taken as an early indication that the mechanism for 
superconductivity is different in the cuprates and is mainly
electronic in nature. However, it was found that in the underdoped region
of the phase diagram the isotope coefficient $\alpha$ increases
rapidly and becomes larger than $1/2$ \cite{franck91,franck94}.
Dahm \cite{dahm00} provided a brief review of some of many possible
explanations including the idea that important energy dependence in
the electronic density of states (EDoS) at the Fermi level
\cite{schach90} on the scale of the superconducting energy can
affect the isotope effect. Zeyher and Greco \cite{zeyher09}
used in comparison to Dahm a far more involved approach but also
arrived at the conclusion that only a small contribution of
electron-phonon interaction to the over all pairing potential
is sufficient to explain the isotope effect in the underdoped
cuprates.

Prominent kink like structures in the electronic dispersion curves
of the cuprates measured in angular resolved photo-emission
spectroscopy (ARPES) \cite{bogdanov00,lanzara01} have been widely
interpreted as indication of the coupling of the charge carriers
to boson modes. Some argued for phonons \cite{bogdanov00,lanzara01}
while other authors favored coupling to a spin fluctuation
resonant mode \cite{johnson01}. Additional information on the origin
of the nodal direction dispersion kinks at $\sim 70\,$meV was
provided by Iwasawa \textit{et al.} \cite{iwasawa08} who measured their
change in optimally doped Bi$_2$Sr$_2$CaCu$_2$O$_{8+\delta}$ on
oxygen substitution $^{16}$O to $^{18}$O. A recent scanning tunneling
spectroscopy (STM) study \cite{lee06} has also found a shift in a mode
at $52\,$meV which is compatible with the expected amount on
$^{16}$O to $^{18}$O substitution. A detailed analysis of the data
on oxygen substitution was undertaken by Schachinger \textit{et al.}
\cite{schach09a,schach10}. Within an Eliashberg formulation of a
boson exchange mechanism one can invert ARPES  \cite{schach09b} or
infrared optical data (IR) \cite{yang09a,yang09b} to recover the 
entire electron boson spectral density $I^2\chi(\omega)$ involved.
This is a composite of phonon structures as well as spin fluctuation
exchange processes or other excitations that could be coupled to the
charge carriers and manifest in the electron boson spectral density
$I^2\chi(\omega)$ \cite{carb11}. Using such an information
Schachinger \textit{et al.} \cite{schach09a} concluded that the 
substitution data of Iwasawa \textit{et al.} \cite{iwasawa08}
can be understood as arising from a phonon peak in $I^2\chi(\omega)$
which accounted for about 10\% of the total spectral weight associated
with the electron boson spectral density. Furthermore, Dal Conte
\textit{et al.} \cite{DalConte12} reported a phonon contribution of
approx. 20\% to the electron boson spectral density in the system
Bi$_2$Sr$_2$Ca$_{0.92}$Y$_{0.08}$Cu$_2$O$_{8+\delta}$ using a very
different type of analysis. All this puts a serious constraint on
the resulting isotope effect and is compatible with the very small 
value of $\alpha\simeq 0.05$ found in the optimally doped cuprates.
The question then arises of how this observation is compatible
with the very large increase in $\alpha$ found as doping is reduced.
One should keep in mind that the properties of the superconducting
state are also observed to evolve away from a BCS description
with $d$-wave superconducting gap symmetry even though such a model
provides a good qualitative description of the optimum and overdoped
region of the cuprate phase diagram.
The observed deviations go beyond what can be understood when
anisotropy \cite{leavens71,leung76,donovan95a,donovan95b,donovan95c,branch95},
energy dependence in the EDoS \cite{mitrovic83a,mitrovic83b} or when
inelastic and strong coupling corrections \cite{nicol91,carb95,hwang06}
are introduced in a generalized Eliashberg description. Many of these
anomalous properties can, however, be understood within the semi
phenomenological model of Yang, Rice, and Zhang (YRZ)
\cite{yang06,konik06,yang09c} of the pseudogap state of the underlying normal
state \cite{timusk99,norman05} of the underdoped cuprates which do
not behave like Fermi liquids would. This model is based on ideas of
the resonant valence bond (RVB) \cite{anderson87} spin liquid and provides a specific
ansatz for the normal state self energy. When generalized to include
superconductivity it has been remarkably successful in providing a
first understanding of the, until then, anomalous properties
\cite{timusk99,norman05} associated with the formation of a pseudogap.
This is accompanied with Fermi surface (FS) reconstruction from the
large FS of Fermi liquid theory to closed Luttinger pockets centered
about the nodal direction in the copper-oxygen plane Brillouin zone (BZ).
As half filling is approached the pockets become progressively smaller. In this
model metallicity is lost not through an ever increasing effective
mass. Rather the quasiparticles remain light in the nodal direction
while the antinodal regions \cite{leblanc11} are completely gaped out.
As a consequence, the in-plane resistivity remains metallic \cite{ashby13}
as observed \cite{ito93,takenaka94}. For example, Lee \textit{et al.} \cite{lee05}
found that a Drude response remains well inside the antiferromagnetic
dome seen at small doping beyond the end of the superconducting dome.
At the same time the model can explain
an insulating $c$-axis dc-response \cite{ito93,takenaka94} within a
coherent $c$-axis charge transfer model which eliminates $c$-axis transport
because the plane to plane matrix element has $d$-wave symmetry and
is zero in the nodal directions \cite{levchenko10,xiang96}.
Among the many results obtained in the literature we mention here
a representative set which shows that the YRZ model is capable of
providing a first understanding of most of the anomalous superconducting
properties seen in the underdoped cuprates. A review can be found in
the paper by Rice \textit{et al.} \cite{rice12} which also provides
elaboration of the theoretical basis of the YRZ ansatz.

Yang \textit{et al.} \cite{yang09c,yang08} showed that the particle-hole asymmetry
observed in ARPES as one moves off the nodal direction in the
pseudogap state can be understood both in terms of their energy
value and their quasi particle spectral weight. LeBlanc \textit{et al.}
\cite{leblanc11a} also discussed the effect of a particle-hole
asymmetric pseudogap on the Bogliubov quasiparticles and found
reasonable agreement with ARPES data of Hashimoto \textit{et al.}
\cite{hashimoto10}. Yang \textit{et al.} \cite{yang11b} showed that
there exists fully enclosed hole Luttinger pockets and that the 
spectral weight of the FS on the antiferromagnetic side of the
BZ is small as compared with the spectral weight
on the other face of the pocket oriented towards the $\Gamma$-point.
Both shape and area enclosed in the pockets are in agreement with
the YRZ model. STM can also be used to get detailed FS data as
found in the work of Kohsaka \textit{et al.} \cite{kohsaka08} for
example. The observed interference
patterns due to impurity scattering are analyzed to yield
FS contours and, possibly, details of the pseudogap Dirac point
\cite{fisher11a}. Bascones and Valenzuela \cite{basc08} addressed
the issue of checkerboard patterns observed in STM. The total
EDoS can also be obtained from STM data.
Borne \textit{et al.} \cite{borne10}
provided an analysis of the STM data of Pushp
\textit{et al.} \cite{pushp09} and found good agreement with YRZ,
confirming an earlier result of Yang \textit{et al.} \cite{yang09c}
who compared with STM data of Kohsaka \textit{et al.} \cite{kohsaka08}.
The low temperature $(T)$
specific heat \cite{loram04,leblanc09,borne10a} was found to remain
linear in $T$ deep inside the underdoped regime. This observation
provides clear evidence that the quasiparticles around the nodes
which is the only region sampled at low $T$ remain BCS like. At the
same time the specific heat is strongly suppressed below its BCS value as
$T$ is increased towards $T_c$ because now the pseudogap region 
of the BZ zone is sampled. A similar situation holds for the
in-plane penetration depth which remains linear in $T$ at low $T$
but becomes suppressed over its optimal value as $T$ is increased
towards $T_c$. In fact both, experiment \cite{huttema09} and theory
\cite{carb10} find a quasilinear in $T$ dependence over the entire
interval from $T=0$ to $T=T_c$. The behavior of the $c$-axis
penetration depth can also be understood in YRZ \cite{carb13}.
Other studies involve the in-plane ac optical conductivity 
\cite{illes09,pound11,hwang07}, the optical scattering rates \cite{bhalla14}
as well as the $c$-axis optical response \cite{nobrega11,ashby13a,homes93}
for both coherent and incoherent charge transfer in the direction
perpendicular to the copper-oxide planes. The $c$-axis
sum rule, long not understood, also follows in YRZ theory \cite{carb12}.
In a Fermi liquid the optical spectral weight under the real part 
of the optical conductivity is unchanged when the superconducting
state is entered. For the underdoped cuprates measurements
\cite{basov99}
of the $c$-axis optical weight found a serious violation of this
sum rule. The Raman response
\cite{basc08,valenzuela07,leblanc10,tacon06,guyard08}
of the underdoped cuprates is particularly interesting.
The peak position in the B2g response which preferentially
samples the nodal response, is found to decrease with decreasing doping while
the B1g response which samples mainly the antinodal region of the
BZ shows the opposite trend, i.e. increases in energy. This is in
agreement with a superconducting gap which decreases while the 
pseudogap increases, a well documented phenomenon \cite{hufner08}
in the cuprate phase diagram. A comparison of YRZ predictions and experimental
data of the Andreev and single particle tunneling
spectra by Yang \textit{et al.} \cite{yang10} provides further
support for the two gap model with distinct superconducting and
RVB gap. The pseudogap also
modifies the universal limits \cite{lee93,durst00} of the optical
and thermal conductivity \cite{sutherland03,carb11a}. Impurity scattering,
again, drops out in this limit but Gutzwiller factors associated
with the effects of strong correlations
appear. These are not part of a Fermi liquid description.
A detailed comparison of the dynamical spin susceptibility calculated
in YRZ by James \textit{et al.} \cite{james12} with both inelastic
neutron and resonant X-ray scattering data found satisfactory
qualitative agreement at all energy scales considered including the
known hourglass pattern for the magnetic susceptibility.

All of the above described properties have aspects that cannot be understood within
a conventional $d$-wave BCS framework but find a natural explanation
when the YRZ model is used to describe the underlying normal state
with the underlying emergence of a pseudogap. With this success 
highlighted, it is important to know whether or not the observed large
increase in isotope effect with decreased doping towards the 
Mott insulating state can also be explained to be a result of
pseudogap formation. A very simple model of pseudogap formation
on the isotope coefficient $\alpha$ has already appeared \cite{dahm00}
and it has been found that $\alpha$ increases steadily as the 
magnitude of the pseudogap is cranked up without changing the 
amount of the pairing potential that is taken to be due to phonons
as compared to the dominating part coming from an electronic
mechanism. This provided further motivation to apply to this
problem the more realistic YRZ model which includes FS reconstruction
into Luttinger pockets. The parameters of the band structure and
pseudogap used in the original YRZ paper will not be altered as
it lead to good agreement with a large data set. The
observation that only about 10\% of the pairing interaction
in the electron boson spectral density can come from phonons
will be respected and represents a constraint on the work.

In Sec.~\ref{sec:2} we summarize the results associated with
FS reconstruction brought about by the emergence of a finite
pseudogap below a quantum critical point (QCP) at doping
$x=x_c$. The
superconducting state with gap of $d$-wave symmetry is considered
in Sec.~\ref{sec:3} where we also present results for the 
isotope effect. Finally, Sec.~\ref{sec:4} contains a discussion 
and our conclusions.

\section{Formalism Normal State}
\label{sec:2}

\begin{figure}[tp]
  \includegraphics[width=120mm]{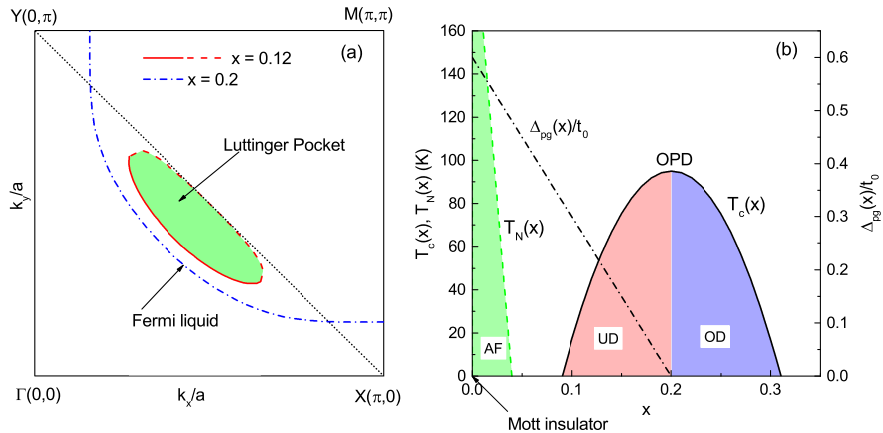}
\caption{(Color online)(a) The right hand upper quadrant of the
Cu$_2$O BZ with momentum \textbf{k} in units of the lattice
parameter $a$. The dashed dotted (blue) curve is the Fermi liquid
large FS at optimum doping $(x=0.2)$ where there is no pseudogap.
The short dotted (black) diagonal represents the AFBZ boundary.
The solid and dashed (red) contour is the FS for the case
$x=0.12$ where the pseudogap has lead to its reconstruction
into a small Luttinger pocket [shaded (green) area] centered
about the diagonal $\overline{\Gamma\textrm{M}}$. (b) A
schematic view of the cuprate phase diagram. The dashed dotted
(black) line (right hand scale applies) shows the pseudogap
amplitude normalized to $t_0$, $\Delta_{pg}(x)/t_0 = 3(0.2-x)$, as a function
of doping $x$ in the YRZ model. The solid (black) dome gives the
superconducting critical temperature $T_c(x)$ vs $x$ (left hand
scale applies) with optimal doping (OPD) indicated as the dome
maximum with the overdoped (OD) [shaded (blue)] area on the right
and the underdoped (UD) [shaded (red)] area on the left. We also
indicated the antiferromagnetic (AF) region [shaded (green) area]
with the N\'eel temperature $T_N(x)$ and
the Mott insulating state at $x=0$.
}
\label{fig:1}       % Give a unique label
\end{figure}

In the YRZ model the coherent part of the charge carrier Green's
function for doping $x$ has the form \cite{yang09c}
\begin{equation}
 \label{eq:1}
% G({\bf k},\omega;x) = \frac{g_t(x)}{\omega-\varepsilon_{\bf k}(x)-
%   \frac{\Delta^2_{pg}({\bf k},x)}{\omega+\varepsilon^\circ_{\bf k}(x)}},
  G^N({\bf k},\omega;x) = \sum\limits_{\alpha=\pm}
  \frac{g_t(x)W^\alpha_{\bf k}(x)}{\omega-E^\alpha_{\bf k}(x)},
\end{equation}
where $g_t(x) = 2x/(1+x)$ is a Gutzwiller factor that enters the 
theory of strongly correlated electrons and relates to the exclusion
of double occupancy because of the strong on-site Hubbard $U$.
If the pseudogap $\Delta_{pg}({\bf k},x)$ is finite there are two branches
$(+/-)$ to the normal state energies, namely
\begin{equation}
 \label{eq:2}
 E^\pm_{\bf k}(x) = \frac{\varepsilon_{\bf k}(x)-\varepsilon^\circ_{\bf k}(x)}{2}
 \pm\sqrt{\Delta^2_{pg}({\bf k},x)+
 \{[\varepsilon_{\bf k}(x)+\varepsilon^\circ_{\bf k}(x)]/2]\}^2} 
\end{equation}
and these are further weighted by factors
\begin{equation}
 \label{eq:2a}
 W^\pm_{\bf k}(x) = \frac{1}{2}\left\{ 1\pm
  \frac{\varepsilon_{\bf k}(x)+\varepsilon^\circ_{\bf k}(x)}
  {2\sqrt{\left[\frac{\varepsilon_{\bf k}(x)+\varepsilon^\circ_{\bf k}(x)}{2}
  \right]^2+\Delta^2_{pg}({\bf k},x)}}
  \right\}.
\end{equation}
The electron dispersion curves $\varepsilon_{\bf k}(x)$ as a function
of momentum \textbf{k} describe the electronic structure when no
account is taken of the pseudogap. It involves up to third nearest
neighbor hoping as well as the magnetic energy scale $J$ of the
$t-J$ model and an additional Gutzwiller factor $g_s(x) = 1/(1+x)^2$.
Details of these dispersion curves are found in the original paper
of YRZ \cite{yang06} and will not be altered here. The energy
$\varepsilon^\circ_{\bf k}(x)$ involves only first nearest neighbor
hopping $t_0$ and $\varepsilon^\circ_{\bf k}(x) = 0$ gives the 
boundary of the antiferromagnetic Brilloin zone (AFBZ). The AFBZ
corresponds to half filling $(x=0)$ which would be metallic if the 
pseudogap was not accounted for but in reality is the Mott insulating
state. It is important to note that in the limit $\Delta_{pg}({\bf k},x)\to 0$
the Green's function $G^N({\bf k},\omega;x)$ reduces to its
usual form, namely $(\omega-\varepsilon_{\bf k})^{-1}$.
 The pseudogap $\Delta_{pg}({\bf k},x)$ is taken to have
$d$-wave symmetry in the BZ with its amplitude linear in $x$,
increasing with decreasing doping and zero at optimum doping where
the critical temperature for superconductivity has its maximum.
At optimum doping taken to be $x_c = 0.2$ in YRZ there is no
pseudogap and the FS is the large contour of Fermi liquid theory
shown as the dash-dotted (blue) line in Fig.~\ref{fig:1}(a)
where the top right quadrant of the CuO$_2$ BZ is shown as a
function of $k_x/a$ and $k_y/a$ with $a$ the lattice parameter.
The short dotted (black) line on the diagonal is the AFBZ. Also
shown is the 
reconstructed FS [solid and dashed (red) line] when $x=0.12$ and
there is
a finite pseudogap. The enclosed area is the
Luttinger hole pocket which we have shaded (green) for emphasis.
The back side of the  Luttinger pocket [dashed (red) line] has small
weighting as compared to the side pointing towards [solid (red) line]
the $\Gamma$-point in the BZ leading
approximately to the concept of a Fermi arc \cite{kanigel07}. 
As the doping is further reduced towards the 
Mott insulating state, the Luttinger pocket shrinks even more.
Nevertheless, a small number of quasiparticles remains along the nodal
direction $\overline{\Gamma\textrm{M}}$. It is in this way
that the metallicity is reduced and eventually lost.

\section{Formalism Superconducting State}
\label{sec:3}

\begin{figure}[tp]
  \includegraphics[width=90mm]{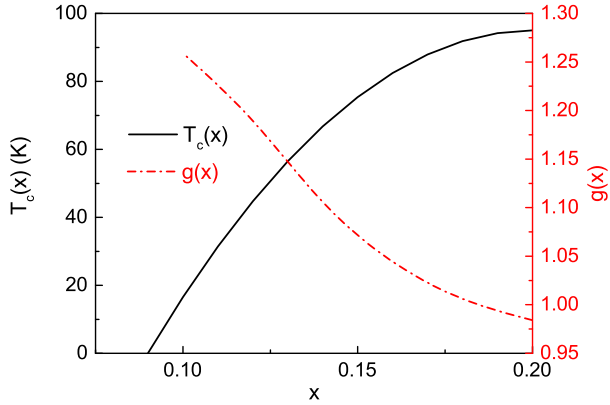}
\caption{(Color online) The coupling constant $g(x)$ [dashed
dotted (red) line, right hand scale] as a function of doping
$x$ which is needed in the nearest neighbor pairing model
[Eq.~\eqref{eq:6}] to reproduce the measured value of $T_c$
as a function of $x$ [solid (black) curve, left hand scale]
in the underdoped region of the cuprate phase diagram
Fig.~\ref{fig:1}(b).
}
\label{fig:2}       % Give a unique label
\end{figure}

We want to build up a superconducting state based on the normal state
which describes the underdoped region of the cuprate phase diagram rather
than on an underlying Fermi liquid band structure described by the
dispersion curves $\varepsilon_{\bf k}(x)$. For a
given pairing potential $V_{{\bf k},{\bf k'}}$ one can then proceed
to write down the superconducting Green's function in the usual way
\begin{equation}
 \label{eq:3}
 G^s({\bf k},\omega,x) = \sum\limits_{\alpha=\pm}\frac{g_t(x)W^\alpha_{\bf k}(x)}
 {\omega-E^\alpha_{{\bf k},s}(x)-\frac{\Delta^2_{sc}({\bf k},x)}{\omega+
 E^\alpha_{{\bf k},s}(x)}},
\end{equation}
where the new energies $E^\pm_{{\bf k},s} \equiv
 \sqrt{[E^\pm_{\bf k}(x)]^2+\Delta^2_{sc}({\bf k},x)}$.
The gap equation at temperature $T$ takes the form \cite{schach10a}:
\begin{equation}
 \label{eq:4}
 \Delta_{sc}({\bf k},T,x) = -\sum\limits_{{\bf k'},\alpha=\pm} V_{{\bf k},{\bf k'}}
  W^\alpha_{\bf k'}(x)\frac{\Delta_{sc}({\bf k'},T,x)}{2
  E^\alpha_{{\bf k'},s}(T,x)}\textrm{tanh}\left(\frac{E^\alpha_{{\bf k'},s}(T,x)}
  {2T}\right).
\end{equation}
Here we included for the first time explicitly the temperature $T$ in the gap.
As we are here interested only in the critical temperature $T_c$
Eq.~\eqref{eq:4} can be linearized in $\Delta_{sc}({\bf k},T,x)$ to give:
\begin{equation}
 \label{eq:5}
 \Delta_{sc}({\bf k},T,x) = -\sum\limits_{{\bf k'},\alpha=\pm} V_{{\bf k},{\bf k'}}
  W^\alpha_{\bf k'}(x)\frac{\Delta_{sc}({\bf k'},T,x)}{2
  \vert E^\alpha_{{\bf k'}}(T,x)\vert}\textrm{tanh}\left(\frac{E^\alpha_{{\bf k'}}(T,x)}
  {2T_c}\right).
\end{equation}
This equation has the usual form of the normal BCS $T_c$ equation except that
now the pseudogap is built into the energies $E^\alpha_{\bf k'}(x)$ and these
go to zero only on the Luttinger Fermi contours which define the Luttinger pockets
of Fig.~\ref{eq:1}(a). To proceed we need a model for the pairing potential.
Two possible models were studied in Ref.~\cite{schach10a} with very similar results.
Here it will be sufficient to take only one, namely
\begin{equation}
 \label{eq:6}
 V_{{\bf k},{\bf k'}} = -g(x)U_{nn}\{\cos[(k_x-k_{x'})/a]+
  \cos[(k_y-k_{y'})/a]\},
\end{equation}
where the nearest neighbor interaction $U_{nn}$ is set equal to
$75\,$meV which leaves one single parameter $g(x)$. The superconducting
dome shown in the cuprate phase diagram Fig.~\ref{fig:1}(b) [solid (black)
line] is an empirical quantity given by
\begin{equation}
 \label{eq:7}
  T_c(x) = 95.0[1-82.6(x-0.2)^2],
\end{equation}
where $T_c(x)$ is given in Kelvin. Optimum doping $(x=0.2)$ gives a
maximum $T_c$ of $95\,$K which is characteristic of YBCO. To the right
is the overdoped [shaded (blue)] and to the left the underdoped
region [shaded (red)]. Near $x=0$ is an antiferromagnetic region
[shaded (green)] as the Mott insulating state is approached.
Equation~\eqref{eq:6} is used to determine the value of $g(x)$
for any doping below $x=0.2$ which covers the underdoped region
with finite pseudogap and reconstructed FS into Luttinger hole
pockets. The results obtained are summarized in Fig.~\ref{fig:2}
where we plot $T_c$ as a function of doping $x$ [solid (black)
curve, left hand scale] and the resulting value of $g(x)$
[dashed dotted (red) curve, right hand scale]. We see that
to reproduce the measured value of $T_c$, $g(x)$ needs to
increase as doping is reduced towards the Mott insulating
state. This may be an indication that the spin fluctuations
increase as the antiferromagnetic region of the phase diagram,
which falls somewhat below the end of the superconducting
dome, is approached.

\begin{figure}[tp]
  \includegraphics[width=120mm]{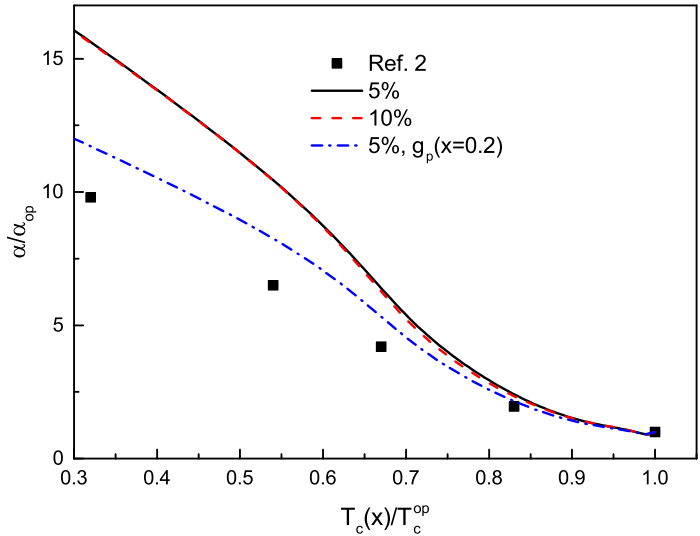}
\caption{(Color online) The isotope effect $\alpha$
normalized to its value $\alpha_{op}$ at optimum doping
as a function of the ratio of the critical temperature
$T_c(x)$ to its value at optimum doping $T^{op}_c$
($x=0.2$ in our model). The solid (black) curve is for a
5\% contribution of phonons to the total pairing coupling
constant $g(x)$ of Eq.~\eqref{eq:6} while the dashed (red)
curve is for a 10\% contribution. The dashed-dotted (blue)
curve leaves the phonon contribution at its value at optimum
doping. The solid black squares are the
data of Ref.~\cite{franck94} for Pr doped YBCO.
}
\label{fig:3}       % Give a unique label
\end{figure}

To discuss the isotope effect at any doping $x$ we break up
$g(x)$ into two contributions. A sub-dominant phonon contribution
which accounts for 5 to 10\% of the total value of $g(x)$ and a
second, dominant piece, electronic in origin, which accounts
for the rest 95 to 90\%. A phonon cut-off at $\omega_D =
80\,$meV is also applied to the phonon contribution and the over
all magnitude of the coupling $g$ is changed to get the measured
value of $T_c$. The phonon cut-off is further
shifted according to the square root of the ratio of the 
oxygen 16 to 18 mass (M) and the calculation of $T_c$ is
repeated, the change in $T_c$ noted and $\alpha$ is then
determined from $T_c\propto M^{-\alpha}$. Results for 
$\alpha(x)$ vs $x$ are presented in Fig.~\ref{fig:3}
where we plot the ratio $\alpha$ to its value at
optimum doping $\alpha_{op}$ as a function of the normalized
temperature $T_c(x)/T_c^{op}$.
Both the case of a 5\% and 10\% phonon contribution to the 
total pairing are presented as the solid (black) and dashed
(red) curves, respectively. We see that $\alpha$
which is small
at optimum doping rapidly increases as $x$ is
decreased into the underdoped region of the phase diagram.
This demonstrates that the existence of a pseudogap can 
drastically increase the isotope effect over its Fermi
liquid value obtained when $\Delta_{pg}(x)$ is set equal to
zero. We also show in Fig.~\ref{fig:3} experimental 
results of Franck \textit{et al.} \cite{franck91,franck94}
obtained for Pr doped YBCO as solid squares. It is clear that
the YRZ model of the underdoped cuprates can naturally provide
an explanation for the anomalous isotope effect observed in
the underdoped cuprates and requires only a 5\% to 10\% contribution
to the pairing potential to originate in the electron-phonon
interaction. This is in line with a large body of other
information on the electron-boson spectral density in the 
cuprates which indicate that the major contribution to the 
pairing glue comes from the exchange of excitations of
electronic origin, probably spin fluctuations \cite{carb11}.
Our results are generic and do not depend sensitively
on details. Rather they have their base in the growth of the
pseudogap which provides an energy dependence to the EDoS.
Because we used a BCS type approach we
cannot expect quantitative agreement with experiment but the
qualitative agreement obtained is robust and the large
increase in $\alpha$ observed is easily understood with a
rather modest magnitude of electron-phonon coupling. The
dashed-dotted (blue) curve serves to emphasis the point
that the exact strength of the electron-phonon interaction
assumed does not change the general trend seen in Fig.~\ref{fig:3}.
To arrive at this curve the electron-phonon part of the 
pairing was kept at its value obtained at optimum doping,
i.e. $x=0.2$. This assumption gives better agreement with
experiment than when $g_p(x)$ is increased slightly with 
decreasing values of $x$ in direct proportion to the 
over all $g(x)$ which, as we saw in Fig.~\ref{fig:2}
[dashed-dotted (red) curve] must increase with decreasing
$x$ so that $T_c(x)$ stays on the measured superconducting
dome [solid (black) curve].

\section{Summary and Conclusion}
\label{sec:4}

The normal state electronic properties of the underdoped
cuprates cannot be understood within a Fermi liquid framework.
Moreover, the observed superconducting state properties also do not
conform, even qualitatively, to the behavior expected in a
BCS model extended to include the $d$-wave symmetry of the 
superconducting gap function. An additional element is
required which goes beyond extensions such as anisotropy,
energy dependent EDoS, inelastic scattering, or other
elaborations which have played some role in the superconductivity
of conventional metals. Strong correlations effects become
essential as the Mott insulating state is approached with reduced doping
and a pseudogap is seen to emerge in the normal state.
How this feature is to be described, however, remains
controversial. A prominent model which has recently shown
great promise is the resonating valence bond spin liquid
model developed by Yang \textit{et al.} \cite{yang06}.
These authors provided a simple phenomenological ansatz for
the self energy in the pseudogap state which has proved very
successful in understanding anomalous normal as well as
superconducting state properties. We provided in this paper a brief
review of these successes to set the context for the present
work. The model involves a
quantum critical point at doping $x=x_c$ below which the 
pseudogap rises in magnitude and strongly modifies the 
electronic structure including the reconstruction of the 
FS from the large contour of Fermi liquid theory to small
Luttinger hole pockets centered about the nodal direction
in the copper-oxygen BZ and near the AFBZ. The energy scale
associated with the pseudogap is comparable to the 
superconducting gap scale and this alters profoundly
superconducting properties as has been documented
here by providing a brief survey of recent 
literature. A conclusion of such work is that, in a large 
part, YRZ can provide a natural and straight forward understanding
of a large variety of properties previously considered anomalous.
Here we extended the work to the isotope effect with equal
success. At optimum doping the observed change in critical
temperature with $^{16}$O $\to\, ^{18}$O substitution is found
to be very small and much less than the BCS prediction for an
electron-phonon system. This observation is consistent with the great deal
of independent knowledge pointing to the fact that the driving
mechanism for superconductivity in the cuprates is mainly electronic
in nature. The evidence \cite{carb11} also points to a
subdominant contribution from phonons which is consistent with
a small, nonzero isotope effect. As the doping is reduced
below optimum doping the pseudogap provides a new energy
dependence to the EDoS and, as we find here, this can increase
radically the  isotope effect. This increase is generic to
such models and requires only a minor contribution to the total
pairing potential to originate from the electron-phonon
interaction. Our calculations are quite
consistent with experimental findings and add another
property of the underdoped cuprates that finds a natural understanding
within the YRZ model.

\begin{acknowledgements}
Research supported in part by the Natural Sciences and Engineering Research
Council of Canada (NSERC) and by the Canadian Institute for Advanced Research
(CIFAR).
\end{acknowledgements}

%\bibliographystyle{spphys}       % APS-like style for physics
%\bibliography{IsoEff}   % name your BibTeX data base

\end{document}